\def \Bbb#1{{#1\kern-.800em #1}}

\def \be {\begin{equation}}
\def \eq {\end{equation}}
\def \bee {\begin{eqnarray}}
\def \eqq {\end{eqnarray}}
\def \n {\nonumber}
\def \bea {\begin{array}{c}}
\def \eqa {\end{array}}

\def \a {\alpha}
\def \b {\beta}
\def \g {\gamma}
\def \d {\delta}
\def \l {\lambda}
\def \lam {\lambda}
\def \del {\partial}
\def \G {{\bf g}}
\def \H {{\bf h}}
\def \K {{\bf k}}
\def \I {{\bf I}}

\def \om {\omega}
\def \omt { {\tilde{\omega} } }

\def \L {{\cal L}}
\def \pl {\psi_L}
\def \pr {\psi_R}
\def \ds {\partial\kern-.5em /}
\def \AA {{\cal A}}
\def \as {{\AA\kern-.5em /}}
\def \V {A_h}
\def \A {A_k}

\documentstyle[12pt]{article}

\begin{document}
\begin{titlepage}
\begin{center}
\today  
\hfill    LBL-38276 \\
\hfill    UCB-PTH-96/05 \\
\hfill    hep-th/9602093\\
\vskip .2in
{\large \bf Non-Abelian Anomalies
and Effective Actions for a Homogeneous Space $G/H$}
\vskip .2in
Chong-Sun Chu,
\footnote{e-mail address: cschu@physics.berkeley.edu}
Pei-Ming Ho
\footnote{e-mail address: pmho@physics.berkeley.edu}
and  Bruno Zumino
\vskip .1in
{\em Theoretical Physics Group\\
    Lawrence Berkeley Laboratory\\
      University of California\\
    Berkeley, California 94720}
\end{center}
\begin{abstract}
We consider the problem of constructing the fully gauged effective action
in $2n$-dimensional space-time
for Nambu-Goldstone bosons valued in a homogeneous space $G/H$,
with the requirement that the action be
a solution of the anomalous Ward identity and be  invariant under
the gauge transformations of $H$.
We show that this can be done whenever the
homotopy group $\pi_{2n}(G/H)$ is trivial, 
$G/H$ is reductive
and $H$ is embedded in $G$ so as to be anomaly free,
in particular if $H$ is an anomaly safe group.
We construct the necessary generalization of the Bardeen counterterm
and give explicit forms for the anomaly and the effective action.
When $G/H$ is a symmetric space the counterterm and the anomaly
decompose into a parity even and a parity odd part.
In this case, for the parity even part of the action, one does not need
the anomaly free embedding of $H$.
                                                      
\end{abstract}
\end{titlepage}

\newpage
\renewcommand{\thepage}{\arabic{page}}
\setcounter{page}{1}

\section{Introduction}
Recently there has been renewed interest \cite{DW,D}
in the study of effective actions which describe the interactions
of Nambu-Goldstone bosons due to the existence of non-Abelian anomalies.
The general case in which the Nambu-Goldstone fields are valued
in a coset space $G/H$ corresponding to a reductive homogeneous space
has been analyzed in the past.
Here $G$ is a Lie group and $H$ a subgroup of it.
In reference \cite{Wu} the action was not fully gauged,
in reference \cite{MSZ} it was but the formulas for the anomaly
and the action were not made totally explicit.
In the present paper we reconsider the question of constructing
the general fully gauged action with the aim of providing
simple derivations of the general formulas and
of making all expressions as explicit as possible
in the case of four space-time dimensions.

In section 2.1 we recall the consistency condition
satisfied by the non-Abelian anomaly and the construction
of the effective action in $2n$-dimensional space-time
from the anomaly \cite{WZ}.
This construction is valid for the general reductive case
when  the anomaly vanishes for the currents of the subgroup $H$,
but for arbitrary external gauge fields in $G$.
In section 2.2 we give  the well known expression
of the Chern-Simon form and of the
 ``canonical anomaly''.    

In general, the  
canonical anomaly for the group $G$ does not have the property
that the currents of the subgroup $H$ are anomaly free,
and therefore cannot be used as it stands to apply the construction
of the effective action described in section 2.1.
If the group $H$ is embedded in $G$ in an anomaly free way,
in particular if $H$ is anomaly safe
\footnote{Anomaly safe groups are groups for which no
representation has anomalies \cite{GG}.
In four dimensions, 
the only simple compact groups which have anomalies
are the $SU(N)$ groups for $N\geq 3$ which includes $Spin(6)=SU(4)$.
$E_6$ is also an anomaly safe group.
}, 
one can find a counterterm to add to the action
which modifies the anomaly so that it has the above mentioned property.
We refer to this as the ``shifted'' anomaly.
This counterterm is a generalization of the counterterm
used by Bardeen \cite{Bar}.
The general form of the counterterm and the expression
for the shifted anomaly are given in section 3.
An explicit form for the counterterm is given for four space-time dimensions.
 
The special case when $G/H$ is a symmetric space is discussed in section 4.
Here there exists a ``parity'' isomorphism of the Lie algebra of $G$
which permits the splitting of the counterterm and of the shifted anomaly
into a parity even and a parity odd part.
It is remarkable that the parity even part satisfies the requirement
of conserved $H$ currents even if $H$ is not embedded
in an anomaly free way.
Both the parity even and the parity odd part can arise from fermion loops,
but only the parity even part seems to be of physical interest.
At the end of section 4, the complete shifted anomaly is given
for the general (nonsymmetric) reductive case in four space-time dimensions.

In Appendix 1, we recall the definition and properties of
the ``homotopy operator'' which was used in sections 2.2 and 3 to obtain
expressions valid in any number of dimensions. The reader who is interested
only in four dimensional space-time can easily check directly all formulas 
using the explicit expressions (\ref{4dstd}), (\ref{B}) 
and following the arguments of section~4.

Finally, in Appendix 2, we give a more geometric description
of the effective action, which follows the ideas developed in references
\cite{Wu,MSZ,Z1,Z2,witten}. It is well known \cite{witten} that the
differential forms which give the anomalies and the effective actions
must be suitably normalized so that the theory can be consistently 
quantized. In the present paper, we shall omit all normalization 
factors and overall factors of $\sqrt{-1}$ in order to write simpler formulas.
These factors can always be reinstated and have been discussed for 
most cases of interest in
\cite{DW,Z1,Z2,witten,ZWZ,BS}.

\section{Non-Abelian Anomaly}

In this section we review some basic facts about the non-Abelian anomaly.

Given a Lie group $G$ and a Lie subgroup $H$, 
we denote their Lie algebras by $\G=Lie(G)$, $\H=Lie(H)$.
We split $\G$ as $\G=\H+\K$;
if $\H$ and $\K$ satisfy
\be
\label{reductive}
   [\H, \K]\subseteq \K,
\eq
then $G/H$ is said to be {\em reductive}.
If further
\be
\label{symmetric}
   [\K, \K]\subseteq \H,
\eq
then $G/H$ is {\em symmetric}.

Notice that $G/H$ admits a reductive splitting whenever $H$ is compact.
The reason is the following.
Given the compact $\H$, we can find its orthogonal complement $\K$
such that $\G=\H+\K$ and $Tr(\l_i\l_a)=0$
in any representation of $\G$ for $\l_i\in\H$ and $\l_a\in\K$.
In general $[\l_i,\l_a]=f_{iaj}\l_j + f_{iab}\l_b$
for $\l_i,\l_j\in\H$ and $\l_a,\l_b\in\K$.
The fact that $f_{iaj}$ vanishes follows from
the cyclic property of trace:
$f_{iaj}=Tr([\l_i,\l_a]\l^j)=Tr([\l^j,\l_i]\l_a)=0$,
where $\l^j$ is the dual of $\l_j$, namely, $Tr(\l_i\l^j)=\d_i^j$.
The dual basis of $\H$ exists because $H$ is compact,
which implies that the Killing metric of $\H$ is not degenerate.

\subsection{Consistency Condition and Effective Action}

In this paper we use Einstein's summation convention
and the convention that
for any $\a=\a^a(x)\l_a$ and $F_a(x)$
\be \label{Einstein}
F_\a = \a\cdot F = \int (d^{D}x) \a^a(x) F_a(x),
\eq
where $D$ could be $2n$ or $2n+1$ for $n=1,2,\cdots$.

In general, the effective action $W[\xi,A]$
of the Nambu-Goldstone boson fields $\xi(x)\in\K$
in the presence of external gauge fields
$A_{\mu}=A_{\mu}^a(x)\l_a$,
where the $\l_a$'s are an anti-Hermitian matrix representation of 
the generators of $\G$,
satisfies the anomalous Ward identity
\be
\label{WG}
   \d_\a W[\xi, A] = \int (d^{2n}x) \a^a(x) G_a[A](x) = G_\a[A],
\eq 
where
$\d_\a = Y_\a + Z_\a$, with $Y_\a$ acting only on $A$
and $Z_\a$ only on $\xi$. 
The operators $Y_\a$ which generate the transformation of the gauge fields are
\be
Y_a(x) = -\del_{\mu}\frac{\d}{\d A_{\mu}^a(x)} -
                f_{abc} A_{\mu}^b(x) \frac{\d}{\d A_{\mu}^c(x)}.
\eq
It is convenient to use the language of differential forms.
Introducing the matrix valued 1-form $A(x)=dx^{\mu}A_{\mu}$, we have
\bee  
  &\d_\a A = d \a +[A,\a],  \label{A-transf}\\
&\d_\a F = [F, \a], \label{transf-F}
\eqq
where $F=dA+A^2$ and
the exterior differential operator $d=dx^{\mu}\del_{\mu}$ commutes with $\d_\a$.

The action of $Z_\a$ on the Nambu-Goldstone bosons which
transform linearly under $H$ and nonlinearly under $G$ \cite{CCWZ}
is specified by the transformation
\be
\label{xi-transf}
   e^{-\xi'} = e^{Z_\a}e^{-\xi} = e^{-\a}e^{-\xi}e^{\eta},
\eq
where $\xi \in \K$ and $\eta\in\H$ is chosen such that $\xi'\in\K$.

An important observation is that the anomaly $G_\a[A]$
must be ``consistent'', i.e. that it must satisfy the 
consistency condition
\be
\label{CC}
   Y_{\a_1} G_{\a_2}[A] - Y_{\a_2} G_{\a_1}[A] = [\a_1, \a_2]\cdot G[A],
\eq
which is a direct result of (\ref{WG}) and the fact that $\d_\a$ 
generates the non-Abelian gauge transformations
and satisfies
\be \label{dd}
\d_{\a_1} \d_{\a_2} -\d_{\a_2} \d_{\a_1} -\d_{[\a_1,\a_2]} =0. \eq
For a more complete explanation see \cite{WZ}.

Now we want to investigate the conditions under which  one can integrate
the anomalous Ward identity to get an effective action $W[\xi, A]$  
satisfying (\ref{WG}).
Given any consistent anomaly $G_\a[A]$,
consider the functional $W[\xi,A]$ as in \cite{WZ}
\be
\label{GW}
   W[\xi,A] = \int_0^1 dt \xi\cdot G[A_t],
\eq
where 
\be
\label{A_t}
A_t = e^{-tY_{\xi}}A = e^{t\xi}Ae^{-t\xi} + e^{t\xi}de^{-t\xi}.
\eq

It is straightforward to check that
\be \label{i1}
   \d_\a A_t = Y_{\a_t} A_t =d \a_t +[A_t, \a_t], \eq
where
\be \label{i2}
 \a_t=e^{t\xi}(\a+Z_\a)e^{-t\xi},
\eq
\be \label{i3}
   \frac{\del}{\del t}A_t = -Y_{\xi} A_t =-(d\xi+[A_t,\xi])
\eq
and
\be \label{i4}
 \frac{\del}{\del t}\a_t = -[\a_t,\xi]- Z_\a \xi. \eq
We have assumed that the action of $Z_\a$ on $\xi$ implied by (\ref{xi-transf})
is well defined, which is the case if $\xi$ is small.

It follows from (\ref{CC}) and (\ref{i1},\ref{i2},\ref{i3},\ref{i4}) that
\bee
\d_\a W 
&=&\int_0^1 dt[(\d_\a \xi)\cdot G[A_t]+\xi\cdot(\d_\a G[A_t])]\n\\
&=&\int_0^1 dt[(\d_\a \xi)\cdot G[A_t]+\xi\cdot(Y_{\a_t}G[A_t])]\n\\
&=&\int_0^1 dt[(Z_\a \xi)\cdot G[A_t]+
                         [\a_t,\xi]\cdot G[A_t]+\a_t\cdot Y_{\xi}G[A_t]]\n\\
&=&\int_0^1 dt[-(\frac{\del}{\del t}\a_t)\cdot G[A_t]
                         +\a_t\cdot(-\frac{\del}{\del t}G[A_t])]\n\\
&=&-\int_0^1 dt \frac{\del}{\del t}(\a_t\cdot G[A_t])\n\\
&=&\a_0\cdot G[A_0] - \a_1 \cdot G[A_1],
\eqq
where $\a_0 = \a$, $A_0 = A$
and $\a_1 = e^{\xi}(\a+Z_\a)e^{-\xi}$.
Note that $\a_1$ belongs to $\H$
since (\ref{xi-transf}) implies that
\be \label{grho}
Z_\a e^{-\xi} = -\a e^{-\xi} + e^{-\xi}\rho
\eq
for some $\rho\in\H$.
Therefore if $G_\a[A]$ also satisfies
\be
\label{G_h}
   G_\a[A] = 0, \quad \a \in \H,
\eq
then the term $\a_1 \cdot G[A_1]$ vanishes and 
$W[\xi, A]$ generates the anomaly as in (\ref{WG}).

This $W[\xi, A]$ is invariant under the gauge transformations 
of $H$. Its expression (\ref{GW}) is very convenient for computing the 
vertices involving different numbers of factors $\xi$.
For example, the Bardeen anomaly \cite{Bar}
is considered  in \cite{WZ} and the corresponding effective action is obtained
by using (\ref{GW}).
It contains the five-pseudoscalar interaction in four
space-time dimensions,
\be \frac{2}{15 \pi^2 F_\pi^5} \int Tr(\Pi (d\Pi)^4), \eq
where $\xi =\Pi/ F_\pi$.
Note that there is a misprint on p.97 of \cite{WZ},
the numerical factor of 1/6 there should be replaced by 2/15.

\subsection{Chern-Simon Form and Non-Abelian Anomaly}
We consider a $2n$-dimensional space-time. In order
to make sense of many formulas given below, we must embed
space-time in a manifold with dimension larger than $2n+1$.
All quantities on space-time such as the gauge field $A(x)$
and the group element $g$
are accordingly extended to the higher dimensional manifold.
Similarly the exterior differentiation $d$ operates in the
higher dimensional manifold.
 
The Chern-Simon form $\om_{2n+1}(A), n=1,2,...$, satisfies
\be \label{CS1}
    Tr F^{n+1} =d \om_{2n+1}(A).
\eq
(The existence of $\om_{2n+1}(A)$ is guaranteed because
$d(Tr F^{n+1})=0$ due to the Bianchi identity $dF=[F,A]$.)
Its form can be obtained, for instance, by using the Cartan homotopy 
operator $k$ \cite{MSZ,Z1} (see Appendix 1) and the one parameter
family of gauge fields $A_t =tA, F_t =dA_t+A_t^2=tF+(t^2-t)A^2$.
Explicitly, it is\cite{MSZ,Z1,Z2} 
\be \label{CS}
\om_{2n+1}(A) = (n+1) \int_0^1 dt Tr(A F_t^n). \eq 
It is easy to see from (\ref{transf-F})
 that the Chern character $Tr F^{n+1}$ is gauge invariant, 
\be \d_\a Tr F^{n+1}=0, \eq
so there exists some $\om^1_{2n}(\a;A)$ such that
\footnote{
Following standard notation the subscript denotes the degree
of the differential form in $dx$ and the superscript its degree in $\a$.
Note that the definitions (\ref{CS1}) 
and (\ref{descent}) do not fix the $\om_{2n+1}$ and
$\om^1_{2n}$ completely: (\ref{CS1}) 
determines the form $\om_{2n+1}$ up to 
$d$ of something while (\ref{descent})
determines  $\om^1_{2n}$ up to $d$ of something plus
$\d_\a$ of something.
}
\be \label{descent}
\d_\a \om_{2n+1}(A) =
   d\om^1_{2n}(\a;A).
\eq
It is well known\cite{Z1} that one can take
\be \label{Std}
 \om^1_{2n}(\a;A) = n(n+1) \int_0^1 dt
 (1-t) Tr(\a d(A F_t^{n-1})).
\eq

For example, the explicit expressions of the $\om$'s
of interest for  4-dimensional physics are
\bee
   \om_5(A) &=&
     Tr(A(dA)^2 + \frac{3}{2}A^3 dA + \frac{3}{5}A^5),\\ 
   \om^1_4(\a;A) &=&
     Tr(\a d(AdA + \frac{1}{2}A^3)).
\label{4dstd}
\eqq
Both are proportional to
\be
   \frac{1}{2}Tr(\l_a\{\l_b,\l_c\}),
\eq
the symmetric invariant $d$-symbol of $\G$.

It follows from (\ref{dd}) and (\ref{descent}) that there exists some
$\om^2_{2n-1}(\a_1,\a_2;A)$ such that
\be \label{om2_2n-1}
   \d_{\a_1}\om^1_{2n}(\a_2;A)-\d_{\a_2}\om^1_{2n}(\a_1;A)
   -\om^1_{2n}([\a_1,\a_2];A)=d\om^2_{2n-1}(\a_1,\a_2;A).
\eq
Hence, the space-time integral  $\int \om^1_{2n}(\a;A)$ 
satisfies the consistency condition (\ref{CC}).
We shall refer to (\ref{Std}) and (\ref{4dstd})
or their space-time integrals as the canonical anomalies.
They have  a meaning in $2n$-dimensional space-time
which is independent of the extension to higher
dimensions.

The anomaly (\ref{Std}) possesses the global symmetry $G$,
but in general it does not vanish on any nontrivial subgroup $H\subset G$.
If it did, its global symmetry would imply that
it also vanishes on the ideal generated by $\H$.
\footnote{A Lie subalgebra $\I \subset \G$ is an ideal if 
$[\G, \I] \subset \I$; the ideal generated by $\H$ is the smallest 
ideal containing $\H$.}   
Therefore a nontrivial anomaly is possible
only if $\H$ belongs to a proper ideal of $\G$.
Since the global symmetry $G$ is not required in most
physical applications,
in order to save the local $H$ symmetry one must
modify this anomaly by adding global $G$ violating 
counterterms to the effective action.
This is what we will do in the next section.

\section{Counterterm Shifting the Anomaly and Effective Action}

We want to construct an 
effective action $W[\xi,A]$ for $\xi\in\K$
with the property that
\be
\label{dvW}
   \d_\a W=0, \quad \a\in\H.
\eq
If we define $W$ by (\ref{GW}),
then (\ref{dvW}) follows if $G_\a$ satisfies (\ref{G_h}).
Therefore what we need is to find a consistent anomaly
$G_\a[A]$ which satisfies (\ref{G_h}).
This will be done in this section. 

In the following we will assume that
$G/H$ is reductive and that $H \subset G$ is an {\em anomaly free embedding}.
The latter means
\be \label{free}
\om_{2n+1}(A) =0, \quad 
\mbox{for $A$ restricted to \H}. \eq
This is equivalent to the condition that
the $d$-symbol for $\G$ has
\be
\label{d-symbol}
   d_{a_1 a_2 \cdots a_{n+1}}\equiv Str(\lam_{a_1},\lam_{a_2},\cdots,\lam_{a_{n+1}})
   = 0
\eq
if all $\lam_{a_1},\lam_{a_2},\cdots,\lam_{a_{n+1}}\in\H$.
The symmetric trace $Str$ for matrix valued forms $C_i$ is defined by
\be
\label{Str}
   Str(C_1,C_2,\cdots,C_N)=\frac{1}{N!}\sum_{P}(-1)^{f(P)}
   Tr(C_{P_1} C_{P_2} \cdots C_{P_N}),
\eq
with the sum  over all the  permutations $P=(P_1, P_2, \cdots, P_N)$
 of $(1,2, \cdots, N)$ and
$f(P)$ is the number of times the permutation $P$
permutes two odd objects.
In the following 
$Str(C_1,\cdots,C_1,C_2,\cdots,C_2,C_3,\cdots)$,
where $C_1$ and $C_2$ appear $n_1$ and $n_2$ times respectively,
will be abbreviated as 
$Str(C_1^{n_1},C_2^{n_2},C_3,\cdots)$.

Define $\V$ and $\A$ by the splitting
\be
   A = \V + \A, \quad \V\in\H, \quad \A\in\K,
\eq
and write
\be
   \a=\b+\g,
\eq
where $\b$ and $\g$ are the $\H$ and $\K$ parts of $\a$,
respectively.

Since $\G=\H+\K$ is reductive,
$\V$ and $\A$ transform respectively as
\be
\label{\V-transf}
   \d_\b \V = d\b + [\V, \b]
\eq
and
\be
\label{\A-transf}
   \d_\b \A = [\A, \b]
\eq
for $\b\in\H$.

Given any one-parameter
family $A_t$ and $F_t=dA_t+A_t^2$, one can introduce the Cartan homotopy
operator $k$  acting on polynomials in $A$ and $F$ 
such that
\be
   k C(A, F)= \int_0^1 l_t C(A_t, F_t)
\eq
and
\be \label{dk}
   (kd + dk) C(A, F) = C(A_1, F_1)-C(A_0, F_0)
\eq
for any polynomial $C$ in $A, F$.
(See Appendix 1 for the definition of $l_t$ and other 
details.)

Now, let  $A_0, A_1$ be two fixed gauge fields and consider 
the Cartan homotopy operator $k$ defined on 
the family of gauge fields
\be \label{At}
   A_t=t A_1+(1-t)A_0.
\eq
Define
\be \label{B01} B_{2n}(A_0,A_1)= -k \om_{2n+1}(A).
\eq
Using (\ref{CS}), it is
\footnote{
Notice that for 2-dimensions, $B_2(A_0,A_1)=Tr(A_0 A_1)$ and the counterterm
$B_2(A_h,A)$ is zero because $Tr(\V^2)=0$ ($\V$ is odd) and 
$Tr(\V \A)=0$ by orthogonality of $\H$ and $\K$. 
The canonical anomaly $\om^1_2(\a;A)$ 
already satisfies (\ref{G_h}).
}
\be  \label{double}
B_{2n}(A_0,A_1) = n(n+1) \int\int_\Delta d\mu d\lam Str(A_0, A_1, F_{\mu,\lam}^{n-1}),
\eq
where $F_{\mu,\lam}=d A_{\mu,\lam} +A_{\mu,\lam}^2$, 
$A_{\mu,\lam}= \mu A_0+ \lam A_1$ and
the  integration is over a triangle $\Delta$ in the $(\mu,\lam)$ plane with
vertices $(0,1), (1,0)$ and the origin.
It is clear from this expression that 
$B_{2n}$ is the generalization of $\a_{2n}$ introduced in \cite{Z1}
\[
  \a_{2n}(dg g^{-1},A)= B_{2n}(-dg g^{-1},A)
\]
for any $g\in G$.
Now we give the derivation of (\ref{double}).
Since, from (\ref{CS}),
\be \om_{2n+1}(A) = (n+1) \int_0^1 dt Tr[A (t F +(t^2-t) A^2)^n], 
\eq
we shall call the parameter in (\ref{At}) $s$ instead of $t$.
Then
\bee 
B_{2n}(A_0,A_1) 
&=& -k \om_{2n+1}(A) \n \\ 
&=& -\int_0^1 l_s \om_{2n+1}(A_s) \n \\
&=& n(n+1)\int_0^1 dt \int_0^1 ds 
    Str(A_s, t \frac{\del A_s}{\del s}, F_{t,s}^{n-1}) \n\\
&=& n(n+1)\int_0^1 dt \int_0^1 ds 
Str(\frac{\del A_{t,s}}{\del t}, \frac{\del A_{t,s}}{\del s}, F_{t,s}^{n-1}), 
\eqq
where $A_{t,s}=t A_s$ and
\bee
F_{t,s}&=&tF_s+(t^2-t)A_s^2=tdA_s+tA_s^2+(t^2-t)A_s^2\n \\
       &=&d(tA_s)+(tA_s)^2= dA_{t,s}+ A_{t,s}^2.
\eqq 
Now changing variables from $(t,s)$ to 
$(\mu, \lam)$ by
\be \lam= t s, \quad \mu= t (1-s), \eq
the integration goes over the triangle $\Delta$.
Notice that
\bee &&A_{t,s}= A_{\mu,\lam}  = \mu A_0+ \lam A_1 , \n \\
     &&dtds\frac{\del A_{t,s}}{\del t}\frac{\del A_{t,s}}{\del s}=
d\mu d\lam\frac{\del A_{\mu,\lam}}{\del \mu}\frac{\del A_{\mu,\lam}}{\del \lam}  
\eqq
and
\be \frac{\del A_{\mu,\lam}}{\del \mu}= A_0, \quad 
\frac{\del A_{\mu,\lam}}{\del \lam}= A_1.
\eq
This completes the proof of (\ref{double}).

For simplicity, the space-time integration considered below 
will be over the compactified $2n$-dimensional
Euclidean space-time $S^{2n}$. 
Define
\be \label{G_v}
G_\a= \int_{S^{2n}} \omt^1_{2n}(\a; \V, A), \eq
where
\be \label{shifted}
\omt^1_{2n}(\a; \V, A)=\om^1_{2n}(\a;A) + \d_\a B_{2n}(\V,A), \eq
i.e.
\be \label{shifted2} 
G_\a= G_{0\a} + \d_\a \int_{S^{2n}} B_{2n}(\V,A), \eq
where
\be
\label{a}
   G_{0\a}[A]= \int_{S^{2n}}\om^1_{2n}(\a;A). 
\eq
We claim that
\be
\label{claim}
G_{\b} =0,\quad \b\in\H.
\eq
First, in view of (\ref{A-transf}) and  (\ref{\V-transf}), we have
 for $A_0=\V$ and $A_1=A$,
\be \label{A01}
\d_\b A_i= d\b+[A_i,\b],  \quad i=0,1,
\eq
so
\be 
\d_\b A_t= d\b+[A_t,\b] \n
\eq
for $A_t=tA_1+(1-t)A_0.$
It is not hard to see that 
\be l_t \d_\b = \d_\b l_t. \eq
Indeed 
\bee l_t \d_\b A_t &=& \d_\b l_t A_t =0,  \n \\
 l_t \d_\b F_t &=& l_t (F_t \b-\b F_t) \n \\
                &=& dt [A_1-A_0, \b] \n
\eqq
and 
\bee
     \d_\b l_t F_t &=& dt \d_\b (A_1-A_0) \n \\
                  &=& dt [A_1-A_0,\b]. \n
\eqq
Therefore
\be k \d_\b =\d_\b k \eq
and so
\bee \d_\b B_{2n}(\V,A)
 &=& -k \d_\b \om_{2n+1}(A) \n \\
 &=& -kd\om^1_{2n}(\b;A) \n \\
 &=& -(kd+dk) \om^1_{2n}(\b;A) +dk \om^1_{2n}(\b;A) \n \\
 &=& -\om^1_{2n}(\b;A)+\om^1_{2n}(\b;\V)+ d k \om^1_{2n}(\b;A). \label{beta}
\eqq
The first term cancels the canonical anomaly,
the second term is zero due to (\ref{d-symbol})
and the third integrates to zero. 
Hence we have proved (\ref{claim}).

For four dimensions, $n=2$, (\ref{double}) gives
\be
\label{B}
   B_4(\V,A)=\frac{1}{2}Tr[(\V A -A\V)(F+F'_h)
                       +A \V^3- \V A^3+\frac{1}{2}\V A \V A],
\eq
where $F'_h=d\V+\V^2$ and $F=dA+A^2$.
It is very easy, using (\ref{A01}), to check 
directly that (\ref{B}) satisfies (\ref{beta}).
Notice that all commutator terms from (\ref{A01}) will cancel under the
trace, so it is sufficient to keep the terms with $d\b$, which makes 
the computation very easy. 

To get an explicit formula for $\omt^1_{2n}(\g; \V, A), \g \in \K$, 
introduce 
\be
\label{omt2n+1}
  \omt_{2n+1}(A_0,A_1) = (n+1) \int_0^1 dt Tr((A_1-A_0) F_t^n)
\eq
for $ A_t=tA_1+(1-t)A_0.$ Using (\ref{dk}) and (\ref{B01}),
it is not hard to check that
\be
\label{omt-B}
\omt_{2n+1}(A_0,A_1) = \om_{2n+1}(A_1) - \om_{2n+1}(A_0)+d B_{2n}(A_0,A_1).
\eq
In fact
\bee 
\lefteqn{\om_{2n+1}(A_1)- \om_{2n+1}(A_0) +d B_{2n}(A_0,A_1)} \n\\
      &=& \om_{2n+1}(A_1) - \om_{2n+1}(A_0) -dk\om_{2n+1}(A) \n\\
      &=& \om_{2n+1}(A_1)-\om_{2n+1}(A_0)-(kd+dk)\om_{2n+1}(A)+ kd\om_{2n+1}(A)
          \n\\
      &=& k Tr F^{n+1} \n\\
      &=& (n+1) \int_0^1 dt Tr((A_1 - A_0) F_t^n).
\eqq
In particular, we have 
\be
\label{om-B}
\omt_{2n+1}(\V,A) = \om_{2n+1}(A)+d B_{2n}(\V,A),
\eq
where we have used again (\ref{free}).
It is clear from (\ref{om-B}) that
\be \label{domt}
Tr F^{n+1} = d\omt_{2n+1}(\V,A),\quad F=dA+A^2 \eq
and
\be \label{descent'}
\d_\a \omt_{2n+1}(\V,A) = d \omt^1_{2n}(\a;\V,A). \eq
Now we can extract $\omt^1_{2n}$ from (\ref{descent'}) using (\ref{omt2n+1}).
Although (\ref{shifted}) is not of this form, we can choose $\omt^1_{2n}(\g; \V, A)$ 
to be of the form $Tr(\g Q)$ up to exact terms.
With this choice 
it is enough, in order to find $Q$, to collect the $d\g$ terms in $\d_\g \omt_{2n+1}(\V,A)$. 
For $\g \in \K$, it is
\be \label{s1} \d_{\g} A = d\g +[A,\g] \eq
and
\bee
\d_\g \V &=& {[\A,\g]}_h =  [\A,\g] - u, \\
\d_\g \A &=& d\g + [\V,\g] +u, \\
\d_\g F_h&=&[F_k,\g] -du -[\V,u]_+ +[\A,u]_+, \\
\d_\g F_k&=&[F_h,\g] +du+[\V,u]_+ -[\A,u]_+ , 
\eqq
where
\be
u= [\A,\g]_k
\eq
is the $\K$ component of $[\A,\g]$ 
and 
\be \label{s6}
F_k=d\A+\V \A+\A \V, \quad F_h=d\V+\V^2+\A^2. 
\eq
Noticing that $F_t=F_h +t F_k +(t^2-1) A_k^2$, we obtain in agreement 
with \cite{MSZ}, where the result was obtained by a different argument,
\bee \label{shifted'}
\omt^1_{2n}(\g; \V, A)= (n+1) \int_0^1 dt &[&Tr(\g F_t^n) 
+ n (t^2-1) Str(\A, [\A,\g], F_t^{n-1}) + \n \\ 
  &&+n (1-t) Str(\A, u,F_t^{n-1} )].  
\eqq

This is a general formula for the shifted anomaly
in any number of dimensions.
It can be used in (\ref{GW}) to obtain the effective action.
In section 4, instead of using (\ref{shifted'}),
 we shall compute explicitly the shifted anomaly
for four dimensions by using (\ref{shifted}) and (\ref{B}) directly.
This will exhibit an interesting feature of the anomaly
in the case of a symmetric space, see (\ref{om-}) below and the remark
immediately after it.

Finally we point out that $\omt_{2n+1}(A_0, A_1)$ is invariant under
simultaneous gauge transformations of $A_0$ and $A_1$
\be \label{simu}
 \omt_{2n+1}( (A_0)^ g , (A_1)^g ) =\omt_{2n+1}(A_0, A_1)
\eq
for any $g \in G$
where 
\be A^g = g^{-1} (A+d) g , \quad g \in G. \eq
This is clear from its definition (\ref{omt2n+1}).

To get the effective action,
we use (\ref{GW}) to integrate the anomaly (\ref{shifted2}).
Using (\ref{i2}), we have
\bee
\int_0^1 dt \int_{S^{2n}} Y_\xi B_{2n}((A_t)_h,A_t) 
&=& - \int_{S^{2n}}  \int_0^1 dt \frac{\del}{\del t} B_{2n}((A_t)_h,A_t) \n\\
&=& - \int_{S^{2n}}B_{2n}((A^g)_h, A^g)+ \int_{S^{2n}}B_{2n}(A_h, A), \n
\eqq 
where $A_t=e^{t\xi}Ae^{-t\xi}+e^{t\xi}de^{-t\xi}$, $g=e^{-\xi}$
and $A^g=(A^g)_h + (A^g)_k$ is the splitting $(A^g)_h\in\H, (A^g)_k\in\K$.
Therefore, 
\be \label{eff-act}
   W[\xi,A]=\int_0^1 ds \xi\cdot G_{0}[A_s] 
   - \int_{S^{2n}}B_{2n}((A^g)_h, A^g)+ \int_{S^{2n}}B_{2n}(A_h, A).
\eq
Using (\ref{Std}), the first term is
\be \label{expW}
   n(n+1)\int_{S^{2n}}\int_0^1 ds \int_0^1 dt(1-t) Tr(\xi d(A_s F_{t,s}^{n-1})),
\eq
where $F_{t,s}=dA_{t,s}+A_{t,s}^2$ with $A_{t,s}=tA_s$.
As explained at the end of Appendix 2, (\ref{expW}) is equal to
the simpler expression, in agreement with \cite{Z1},
\be
  \int_{S^{2n}} L_0-\int_{S^{2n}}B_{2n}(-dgg^{-1},A),
\eq
where
\be
  L_0=(-1)^{n}\frac{n!(n+1)!}{(2n)!}\int_0^{1}dt Tr(\xi U^{2n}_t),
\eq
$g=e^{-\xi}$ and
\be \label{Ut}
   U_t= e^{t\xi}de^{-t\xi}.
\eq
It is obvious that 
\be \label{dUt}
  U_t^{2n}=-dU_t^{2n-1}.
\eq
When all gauge fields vanish, the surviving effective action
is given by the Lagrangian density $L=L_0-B_{2n}(U_h,U)$,
where $U=g^{-1}dg=U_h+U_k$ for $U_h\in\H, U_k\in\K$.
It can be checked that $L$ changes by an exact term under a global $G$ transformation.
Therefore the action $\int_{S^{2n}}L$ is invariant.

Together with the standard (nonlinear) kinetic energy term \cite{CCWZ}
for $\xi$, $W[\xi,A]$ gives 
the interaction of the Nambu-Goldstone bosons in the presence
of external gauge fields $A$.  
It is perhaps worthwhile to emphasize again that $W[\xi, A]$ is not 
globally $G$-invariant:
the Bardeen counterterm $B_{2n}$ breaks the invariance explicitly.
In the case of $G=SU(3) \times SU(3), H=SU(3)_V$ \cite{WZ},
$W[\xi, A]$ can be used, in the approximation of vector meson
dominance, to estimate the vertices which describe the strong interactions
between vector , axial vector and pseudoscalar mesons. It is the 
shifted form of the anomaly which agrees reasonably well with 
experiment \cite{kay}.

\section{Parity and Effective Actions}

The result for $W[\xi,A]$ in the last section is not the only possible one
with the property that the anomaly $G_\a[A]$ vanishes for $\a\in\H$.
If we have a Lie algebra isomorphism
we can decompose the canonical anomaly $\om^1_{2n}$ and $B_{2n}$
further according to their eigenvalues
under the isomorphism.
For physical applications, very often one has to choose $W[\xi,A]$
to respect all symmetry properties of the physical system.
For example, the effective action of the mesons
should be even under parity
transformation and even under charge conjugation.
Here we consider the generalization of the parity transformation
for symmetric spaces,
and find effective actions with definite parity
in addition to the local $H$ symmetry.

For symmetric spaces (\ref{symmetric}),
one can define the ``parity'' transformation
$P$ that corresponds to an isomorphism of the Lie algebra
\be
\label{auto}
   \H\rightarrow\H, \quad \K\rightarrow -\K,
\eq
so that
\bee
   &&P\V = \V, \n\\
   &&P\A = -\A,\n\\
   &&P\xi = -\xi,\n\\
   &&P\a = P(\b+\g) = \b-\g.
\eqq
It is  $P\d_\a=\d_\a P$ for all $\a\in\G$.
One can split the counterterm $B$
\footnote{Here we denote $B_{2n}$ and $\om^1_{2n}$ as $B$ and $\om$
for simplicity.}
and the canonical anomaly $\om$,
\bee
 B=B_+ + B_-, \quad B_\pm = (1 \pm P)B/2, \\ 
 \om=\om_+ + \om_-, \quad \om_\pm = (1 \pm P)\om/2. 
\eqq
For four dimensions, they are
\footnote{In this section we will not write down the unimportant exact terms
which integrate to zero and will not contribute to the anomaly.
Notice that, since $B$ and $\om$ are $4$-forms,
the space-time integrals of $B_-$ and $\om_-$ are even
and those of $B_+$ and $\om_+$ are odd under the physical parity operation which 
changes the sign of some fields and the space part of $x$ in all fields.}
\be \om_+(\a)=-Tr(d\b f_+ + d\g f_-), \quad \om_-(\a)=-Tr(d\b f_- + d\g f_+), \eq
where
\bee &f_+ =\V d\V+\A d\A + 
      \frac{1}{2}(\V^3+\V \A^2 +\A \V \A+\A^2 \V), \n \\
     &f_- =\V d\A+\A d\V + 
      \frac{1}{2}(\A^3+\A \V^2 + \V \A \V+\V^2 \A) 
\eqq
and
\bee
&B_+= Tr[\frac{1}{2}(\V \A-\A \V)F_k
                       -\frac{1}{4}\V \A \V \A], \\
&B_-= Tr[(\V \A-\A \V)F_h
                       -\frac{3}{2}\V \A^3 -\frac{1}{2}\V^3 \A].
\eqq

For $\a=\b$, it is 
\be \label{om+} \om_+(\b)+\d_\b B_+ =-Tr[d\b(\V d\V +\frac{1}{2} \V^3)] \eq
and
\be  \label{om-} \om_-(\b)+\d_\b B_- =0. \eq
Notice that one can always shift the anomaly $\om_-$ to vanish on $\H$ 
by using the counterterm $B_-$, we don't have to assume that 
$H \subset G$ is an anomaly free embedding.

For $\a=\g$, using (\ref{s1}-\ref{s6}) for $u=0$, one finds
\be \label{B-}
\om_-(\g) +\d_\g B_-= 
Tr[\g(3F_h^2+F_k^2-4(\A^2 F_h+\A F_h \A+F_h \A^2) +8\A^4)], \eq
where $F_h=d\V+\V^2+\A^2$ and $F_k=d\A+\V \A+\A \V$.
This generalizes the result of Bardeen \cite{Bar}.
In his paper he considered essentially $G=U(N)\times U(N)$
for which the trace of parity-even terms vanish
(because $Tr=tr_L - tr_R$) and therefore
$\om_+(\a) + \d_\a B_+=0$ automatically.
For a general symmetric space $G/H$, 
one finds,
\bee \label{B'+}
\om_+(\g) + \d_\g B_+&=&
  \frac{3}{2}Tr[\g (F_h F_k +F_k F_h -F_k \A^2 -\A F_k \A -\A^2 F_k)]\n\\
   && +Tr [ [\A,\g][F_h', \V]_+ ] - \frac{1}{4}Tr [ [\A,\g][\V,\V^2]_+ ]. 
\eqq
The last line vanishes because the $d$-symbol is zero
when restricted  to $\H$ and one obtains 
the following non-trivial anomaly,
\be \label{B+}
\om_+(\g) + \d_\g B_+ =
  \frac{3}{2}Tr[\g (F_h F_k + F_k F_h -F_k \A^2 -\A F_k \A -\A^2 F_k)].
\eq
Using ${G_\a}_{\pm}(A)=\int_{S^{2n}}(\om_\pm(\a; A)+\d_\a B_{\pm})$,
one can get the corresponding parity-definite effective action by (\ref{GW}).
They agree with those obtained by doing
the parity splitting directly on $W[\xi, A]$.  

Notice that both $\om_{\pm}$ satisfy the consistency condition and
both are non-trivial, in the sense that they cannot be written as
the $\d_\a$ of something local.
It is well known that the canonical anomaly (\ref{4dstd}) arises
when Weyl fermions are coupled to the external gauge fields, 
it is natural to ask what are the corresponding fermion actions
that give rise to $\om_\pm$. 
Consider the Lagrangian 
\be 
\L(\psi_1, \psi_2, \V, \A)= i\overline{\psi_1} (\ds+\as_+) \psi_1 
    +i\overline{\psi_2} (\ds +\as_-) \psi_2, 
\eq
where
\be \AA_-=\V-\A, \quad \AA_+=\V+\A. \eq
It is obvious that $\L$ is invariant under the infinitesimal transformation
\bee && \d_\a \psi_1 = (\b+\g) \psi_1, \n\\
     && \d_\a \AA_+ =  d\b+d\g + [\AA_+, \b+\g], \n\\
     && \d_\a \psi_2 = (\b-\g) \psi_2, \n\\
     && \d_\a \AA_- =  d\b-d\g + [\AA_-, \b-\g] 
\eqq
for arbitrary $\b \in \H, \g \in \K$.
One then obtains $ \om_-$ from $\L(\pr, \pl, \V, \A)$ for
a set of left-handed spinors $\pl$ and a set of
right-handed spinors $\pr$ and
$\om_+$ from $\L(\pr, \pr', \V, \A)$ for 2 sets of right-handed
spinors $\pr, \pr'$.

Finally, we give the explicit form of the anomaly which vanishes on $\H$
for the reductive case (\ref{reductive}) in 4 dimensions.
Define $\d' = \d_\g -\d_\g^{symm}$,
where $\d_\g$ acts on the gauge fields according to (\ref{s1})-(\ref{s6}) and
$\d_{\g}^{symm}$ is given by the corresponding formula with $u$ set equal
to zero.
Then
\bee
& \d' \V =-u, \quad \d' A = 0,  \n\\
& \d' F_h = -du -[\V,u]_+ +[\A,u]_+, \quad \d' F=0.
\eqq

Since $\d_{\g}^{symm} B$ has already been given in (\ref{B-}) and (\ref{B'+}),
it is enough to calculate $\d' B$.
It follows from (\ref{B}) straightforwardly
that
\bee \label{B'}
 \d' B =- \frac{1}{2}Tr[u ([F,\A]_+ +2[F'_h, A]_+ -\V^3 -\A^3 )].
\eqq

Adding (\ref{B-}), (\ref{B'+}) and (\ref{B'}), one obtains the full answer
for the shifted anomaly for the reductive case in four space-time dimensions
\bee \label{omt}
\omt=\lefteqn{ \om(\g) + \d_\g B =} \n \\
&&Tr[\g (3F_h^2+F_k^2-4(\A^2 F_h+\A F_h \A+F_h \A^2) +8\A^4)] + \n\\
&&\frac{3}{2}Tr[\g (F_h F_k +F_k F_h -F_k \A^2 -\A F_k \A -\A^2 F_k)]+\n\\
&&- \frac{1}{2}Tr[u ([F+ 2F'_h,\A]_+  -\A^3)].
\eqq
Here we have kept that part in the last line of (\ref{B'+})
which does not vanish when $G/H$ is not symmetric.
This expression (\ref{omt}) is the one to be used in (\ref{G_v}) and (\ref{GW})
for the general reductive case.

\section{Acknowledgement}

We are grateful to Steven Weinberg for 
asking the questions which led to the present investigation and for helpful
remarks.
This work was supported in part by the Director, Office of
Energy Research, Office of High Energy and Nuclear Physics, Division of
High Energy Physics of the U.S. Department of Energy under Contract
DE-AC03-76SF00098 and in part by the National Science Foundation under
grant PHY-9514797.

\section{Appendix 1}

For the sake of completeness, we shall review the definition and
basic properties of the Cartan homotopy operator $k$ in this Appendix.
Consider the graded algebra ${\cal G}$  spanned by the one-form $A$
and the two-form $dA$. Equivalently, one can use the
generators $A$ and $F=dA+A^2$.
The exterior derivative $d$ increases the degree of
a form by 1. We want to introduce a second operator $k$ on ${\cal G}$ so
that it decreases the degree of
a form by 1 and behaves in some sense like ``$d^{-1}$''.
   
For any one-parameter family of gauge fields $A_t$ and $F_t=dA_t+A_t^2$,
define an operator $l_t$ as
\be l_t A_t =0, \quad l_t F_t = d_t A_t = dt \frac{\del A_t}{\del t} \eq
and extend it to an arbitrary polynomial $C(A_t, F_t)$ as a derivation or
anti-derivation depending on whether we take $dt$ to be odd or even.
Here we shall take $dt$ to be odd and so $l_t$ acts as a derivation.
The Cartan homotopy operator $k$ \cite{MSZ,Z1}
is defined on any element of ${\cal G}$
which is a polynomial $C(A,F)$ as
\be k C(A, F)= \int_0^1 l_t C(A_t, F_t) .\eq
We will treat $\int_0^1$ as odd since the combination $\int_0^1 dt$ is
even.
We take the convention that we move $dt$ next to $\int_0^1$ before integrating.

It is easy to check that
\be l_t d -d l_t = d_t \eq
 on  ${\cal G}$.
Indeed,
\bee (l_t d -d l_t )A_t &=& l_t (F_t -A_t^2) = d_t A_t, \n\ \\
     (l_t d -d l_t )F_t &=& l_t (F_t A_t -A_t F_t) -d (d_t A_t) \n \\
                        &=& (d_t A_t) A_t -A_t d_t A_t + d_t (dA_t) \n \\
                        &=& d_t F_t \n
\eqq
and so
\bee (kd +dk)C(A,F) &=& \int_0^1 (l_t d -d l_t ) C(A_t, F_t) \n\\
                    &=& \int_0^1 d_t C(A_t, F_t) \n\\
                     &=& C(A_1, F_1)-C(A_0, F_0). \label{kd+dk}
\eqq
In the above we have used the Bianchi identity
\[ dF_t = F_t A_t - A_t F_t. \]

As an application of (\ref{kd+dk}), (\ref{CS}) is obtained from
(\ref{CS1}) for $A_t=t A$. It is
\be
   d\om_{2n+1}(A)=Tr F^{n+1}=(kd+dk)Tr F^{n+1}=dk Tr F^{n+1}.
\eq
Hence
\be
   \om_{2n+1}(A) = k Tr F^{n+1} = (n+1) \int_0^1 dt Tr(A F_t^n).
\eq 

These considerations can be extended to a multi-parameter family of gauge fields
\cite{MSZ} but we do not need this generalization here.

\section{Appendix 2}

In this Appendix, we describe and extend the results of the main text in more
geometric terms.
The elements of $G/H$ are equivalence classes of elements of $G$. In each class we choose 
an element 
$g \in G$ to represent that class. This determines a subset of $G$ which is 
in one-to-one correspondence with $G/H$. 
If $g$ is sufficiently close to the identity of $G$, it can be 
parametrized as
\be \label{gxi}
g = e^{-\xi}, 
\eq
where $\xi \in \K$ \cite{CCWZ}.
However, most of the formulas of this Appendix have more general
validity.

Denote by $D_{2n+1}$ a $(2n+1)$-dimensional disk
with its boundary fixed by the compactified space-time $S^{2n}$.
We extend the gauge field $A(x)$ and the group element $g(x)$
from $S^{2n}$ to $D_{2n+1}$.
Note that $g$ can always be extended to $D_{2n+1}$ when 
the homotopy group $\pi_{2n}(G/H)$ is trivial, which we shall assume.
The exterior differentiation $d$ operates in $D_{2n+1}$ as well.

Consider the effective action of $g$ in the presence of $A$
\be \label{W[A]}
   W[g,A]=\int_{D_{2n+1}}(\omt_{2n+1}(\V,A)-\omt_{2n+1}((A^g)_h,A^g)),
\eq
where
\be
   A^g =g^{-1}Ag + g^{-1}dg\label{A^g}
\eq
and 
$(A^g)_h \in \H, (A^g)_k \in \K$ is the splitting 
of $A^g  = (A^g)_h+(A^g)_k.$ 
For simplicity, we will denote them as $A^g_h$ and $A^g_k$ in the following. 
It is, in agreement with (\ref{grho}),
\be 
\d_\a g = -\a g +g \rho 
\eq
for a suitable $\rho \in \H$. It is easy to 
check that $A^g$ and $A^g_h$ transform as
\be
\d_\a A^g = d \rho +[A^g, \rho] \n 
\eq
and
\be
\d_\a A^g_h = d \rho +[A^g_h, \rho]. 
\eq
So the term $\omt_{2n+1}(\V^g,A^g)$ is invariant due to (\ref{simu}) 
and, from (\ref{descent'}),  
the variation of $W$ in (\ref{W[A]}) is the shifted anomaly
\be
   \d_\a W[g,A]=\d_\a\int_{D_{2n+1}}\omt_{2n+1}(\V,A)
   =\int_{S^{2n}}\omt^1_{2n}(\V,A).
\eq
Since the  integrand  of (\ref{W[A]}) is exact 
\be
   d(\omt_{2n+1}(\V,A)-\omt_{2n+1}(\V^g,A^g))=TrF^{n+1}-Tr(g^{-1}Fg)^{n+1}=0,
\eq
where (\ref{domt}) was used,
(\ref{W[A]}) depends only on the values of
the integrand on the boundary $S^{2n}$ if $\pi_{2n+1}(G/H)$ is trivial.
Otherwise, the coefficient of the action  gets quantized \cite{witten}.
For a more detailed discussion of special situations see \cite{DW,D}.

A more explicit expression for $W[g,A]$ (\ref{W[A]}) can be obtained
by using (\ref{om-B})
\be
\label{W}
   W[g,A]=
        \int_{D_{2n+1}} (- \omt_{2n+1}(\V^g,A^g) + \om_{2n+1}(A))
        +\int_{S^{2n}} B_{2n}(\V,A).
\eq

Let us consider (\ref{W}) term by term.
The first term is gauge-invariant, but its integrand is in general not closed.
The second term is there so that the sum of the integrands
of the first two terms is closed.
The variation of the second term gives
the canonical anomaly $G_0$ as in (\ref{a}).

Using (\ref{omt-B}), we see that
\bee
&&\omt_{2n+1}(A_h^g,A^g)=\om_{2n+1}(A^g)+dB_{2n}(A_h^g,A^g),\n\\
&&\omt_{2n+1}(-dgg^{-1},A)
=\om_{2n+1}(A)-\om_{2n+1}(-dgg^{-1})+dB_{2n}(-dgg^{-1},A),\n
\eqq
where we have used the anomaly free condition for $\H$.
Noticing that $\om_{2n+1}(-dg g^{-1})=-\om_{2n+1}(g^{-1}dg)$
and $\omt_{2n+1}(-dgg^{-1},A)=\omt_{2n+1}(0,A^g)=\om_{2n+1}(A^g)$
due to (\ref{simu}),
we can combine the above two equations to
rewrite the first two terms in $W[g, A]$ more explicitly as
\bee
   \lefteqn{\int_{D_{2n+1}} ( - \omt_{2n+1}(\V^g,A^g) + \om_{2n+1}(A) )
       = \int_{D_{2n+1}} [-\om_{2n+1}(U)] } \n\\
 && -  \int_{S^{2n}}[B_{2n}(\V^g,A^g) + B_{2n}(-dgg^{-1},A)],\label{b}
\eqq
where $U=g^{-1}dg$.

The Wess-Zumino-Witten term $\L_{WZW}$ \cite{DW,D,Wu,MSZ,WZ,Z1,witten} 
is the only surviving 
term in the effective action
when all gauge fields are set to zero. It is 
\be \label{lwzw}
   \L_{WZW} = -\om_{2n+1}(U) -dB_{2n}(U_h, U),
\eq
where $U=g^{-1} dg =U_h +U_k$ for $U_h \in \H, U_k \in \K$. 
It is easy to see that
\bee
   \L_{WZW} &=& -\omt_{2n+1}(U_h,U)\n\\
           &=& -(C^{2n+1}_{n+1})^{-1} 
               \sum_{j=0}^n (-1)^j 
               {C^{2n+1}_{n-j}} Str (f_h^{n-j}, U_k^{2j+1}), \label{lwzw2}
\eqq
where $C^n_m=\frac{n!}{m!(n-m)!}$ is the binomial coefficient
and $f_h=dU_h+U_h^2$.
For $\L_{WZW}$ to be nonzero, we need dim$(G/H) \geq 2n+1$.  

In particular for 4-dimensional space-time,
one finds, in agreement with \cite{DW},
\be 
   \L_{WZW} = -\frac{1}{10}Tr(U_k^5-5U_k^3 f_h+10U_k f_h^2).
\eq

The integral $\int_{D_{2n+1}}[-\om_{2n+1}(U)]$ in (\ref{b}) can be reduced to
a $2n$-dimensional space-time integral.
It is 
\bee
-\om_{2n+1}(U) &=& (-1)^{n-1}\frac{n! (n+1)!}{(2n+1)!} Tr(U^{2n+1}).
\eqq

As explained above, 
when $g$ is sufficiently close to the identity  of $G$, 
the Nambu-Goldstone
bosons $\xi$ are given by (\ref{gxi}).
A convenient choice of coordinates for $D_{2n+1}$ is $(t,x)$
with $t\in[0,1]$ and $x \in S^{2n}$. The group element becomes
a function $g(t,x)$ with $g(1,x)= e^{-\xi(x)}$ and $g(0,x)=1$.
The differentiation operator $d$ in $2n+1$ dimensions 
becomes now 
\be d=d_t+d_x, \eq
where $d_t =dt \frac{\del}{\del t},
d_x=dx^i \frac{\del}{\del x^i}$.
We have
\bee U&=&g^{-1}dg \n\\
      &=& g^{-1} d_x g  + g^{-1} d_t g\n\\
      &=& U_t +dt g^{-1} \del_t g,
\eqq
where $U_t = g^{-1}(t,x) d_x g(t,x)$.
Since $(dt)^2=0$, 
\be 
TrU^{2n+1} = (2n+1) dt  Tr( g^{-1} \del_t g \; U_t^{2n}).
\eq
A particular convenient extension for the group element is
\be
   g(t,x)=e^{-t\xi(x)}.
\eq
This corresponds to (\ref{GW}).
With this choice,
\be
  -\om_{2n+1}(U) =(-1)^n \frac{n!(n+1)!}{(2n)!} dt Tr(\xi U_t^{2n}),
\eq
where $U_t= e^{t\xi}d_x e^{-t\xi}$.
In (\ref{Ut}) and (\ref{dUt}) we have simply written $d$ for $d_x$.
The advantage of integrating over $D_{2n+1}$ is that $W[\xi, A]$ as given in
(\ref{eff-act}) and (\ref{expW}) can be
written compactly as (\ref{W[A]}) with $g=e^{-\xi}$ 
and be given a geometrical meaning.

The differential forms $\om_{2n+1}(U)$,
or equivalently $\omt_{2n+1}(U_h,U)$ from (\ref{lwzw},\ref{lwzw2}),
are generators of the cohomology groups of $G/H$ \cite{DW,D}.
Suitably normalized, their integrals over cycles in $G/H$
will give integers which are topological invariants
of $G/H$ and can be related to invariants in the
combinatorial topology of $G/H$.
When the anomaly arises in perturbation theory
from fermion loops its normalization should agree with
that determined geometrically. As mentioned in the Introduction,
the correct normalization has been discussed for most cases of interest in
\cite{DW,Z1,Z2,witten,ZWZ,BS}.

\end{document}